\documentclass[twocolumn,showpacs,preprintnumbers,amsmath,amssymb]{revtex4-1}
\usepackage[pdftex]{graphicx}
\usepackage{epsfig}

\usepackage{amssymb}
\usepackage{amsmath}
\usepackage{color}
\usepackage{comment}
\usepackage[inline]{trackchanges}

\begin{document}

\title{Tailoring the Local Density of Optical States and directionality of light emission by symmetry-breaking}
\author{S. Cueff$^1$}
\email{sebastien.cueff@ec-lyon.fr}

\author{F. Dubois$^1$}
\author{MS.R. Huang$^2$}
\author{D. Li$^2$}
\author{X. Letartre$^1$}
\author{R. Zia$^2$}
\author{P. Viktorovitch$^1$}
\author{H. S. Nguyen$^1$}

\affiliation{$^1$Institut des Nanotechnologies de Lyon, INL/CNRS, Universit\'e de Lyon, 36 avenue Guy de Collongue, 69130 Ecully , France}
\affiliation{$^2$School of Engineering and Department of Physics, Brown University, Providence, RI 02912, USA }

\date{\today}
\pacs{}

\begin{abstract}
We present a method to simultaneously engineer the energy-momentum dispersion and the local density of optical states. Using vertical symmetry-breaking in high-contrast gratings, we enable the mixing of modes with different parities, thus producing hybridized modes with controlled dispersion. By tuning geometric parameters, we control the coupling between Bloch modes, leading to flatband, M- and W-shaped dispersion as well as Dirac dispersion. Such a platform opens up a new way to control the direction of emitted photons, and to enhance the spontaneous emission into desired modes. We then experimentally demonstrate that this method can be used to redirect light emission from weak emitters -- defects in Silicon -- to optical modes with adjustable density of states and angle of emission.  
\end{abstract}

\maketitle

\section{Introduction}
Controlling light emission, propagation, and extraction on-chip has been a subject of intense research for decades~\cite{soref2006past,lipson2005guiding}. The emergence of nanophotonics has helped to provide new tools and new technologies to optimize light-matter interaction, control optical dispersion, enhance nonlinear effects, and boost light emission. It has been known since the pioneering work of Purcell that spontaneous emission is not only controlled by the emitter’s intrinsic electronic levels but is also governed by the surrounding optical environment~\cite{purcell}. Following that groundbreaking discovery, multiple ways to modify spontaneous emission were invented, and a wealth of different configurations and geometries to modify the local density of optical states (LDOS) have been demonstrated. Among them, the most common structures are dielectric microcavities and plasmonic nanostructures~\cite{pelton2015}. For example, photonic crystal nanocavities were used to increase light emission and extraction from defects in crystalline silicon or in Erbium-doped silicon-based materials~\cite{pelton2015,lo2011room,zelsmann2003seventy,fujita2008light} as well as to enhance or inhibit spontaneous emission from quantum dots~\cite{lodahl2004controlling,englund2005controlling,fujita2005simultaneous,noda2007spontaneous}. Similarly, metallic gratings exploiting plasmonic resonances were designed to increase the electric field and enhance the LDOS at the emitters’ location~\cite{makarova2010photonic,pelton2015}.  

In both cases, the emphasis is put on the Purcell effect to control the spontaneous emission rate. However, even though accelerating the emission rate is very useful to increase the flow of emitted photons and the potential modulation speed~\cite{tsakmakidis2016large}, these nanostructures do not usually enable control of the direction and, more generally, the mode into which light is emitted. Controlling the directivity of radiation is a key parameter to couple light into a desired channel. This is particularly true for weak or single-photon emitters on-chip whose emitted light needs to be efficiently funneled into specific modes with very low losses. 

Directional emission was previously reported for quantum emitters coupled to plasmonic nanostructures such as dipolar nanoantennas~\cite{taminiau2008optical,curto2010unidirectional}, plasmonic arrays~\cite{lozano2013plasmonics}, and split-ring resonators leveraging multipolar interferences~\cite{hancu2013multipolar}. However, losses associated with the optical absorption in the metal reduce the emission efficiency. To minimize such losses, all-dielectric systems including planar dielectric antennas~\cite{lee2011planar}, Silicon Mie resonators~\cite{cihan2018silicon,staude2013tailoring}, array of III-V nanopillars~\cite{ha2018lasing} have been recently used to control the emission or scattering direction.

In this work, we show how all-dielectric symmetry-broken photonic membranes can be designed to simultaneously control the directionality of light emission and the LDOS. We then show how this concept can be exploited to enhance weak light emission from silicon into arbitrary directions and with engineered local density of optical states. 

\section{Design and theory}
\subsection{Principle}
High-index Contrast Gratings (HCGs) are wavelength-scale one-dimensional photonic membranes having a relatively simple geometry but an intriguing variety of optical properties. For example, they can be designed to have high reflection properties with adjustable resonant wavelength and/or bandwidth. The latter can be adjusted at will to produce very broadband mirrors or high quality-factor resonators. The physics behind these effects lie in the constructive/destructive interference between (at least two) modes of the HCGs~\cite{chang2012high}. This can be controlled through the membrane thickness, grating period, and fill factor. However, these methods do not enable to control the dispersion of the structure.

We recently demonstrated that, by breaking the vertical symmetry of a HCG, we can precisely adjust the energy-momentum dispersion of the structure~\cite{nguyen2018symmetry}. The underlying physics of this effect is illustrated in Fig.~1. Note that symmetric HCGs contain even and odd waveguided eigenmodes that are orthogonal and therefore do not interact with each other. An example of such eigenmodes are shown in Fig.~1a. The orthogonality between the two modes is depicted by the crossing of their photonic dispersion diagrams. By breaking the vertical symmetry (e.g. by etching only a fraction of the grating), modes of different parities overlap and can now couple, producing new hybridized modes in the photonic band diagram. An example of such hybridization is shown in Fig.~1b, in which anti-crossings between the two photonic dispersions lead to a bandgap-opening. This coupling between modes can be finely tuned through adjusting geometric parameters (e.g., etch depth, fill factor, period, etc.) to produce exotic dispersion shapes. This symmetry-breaking concept therefore provides a new way to engineer the energy-momentum dispersions of arbitrary materials (more details available in \cite{nguyen2018symmetry}). 

\begin{figure}[t]
\begin{center}
\includegraphics[width=8.5cm]{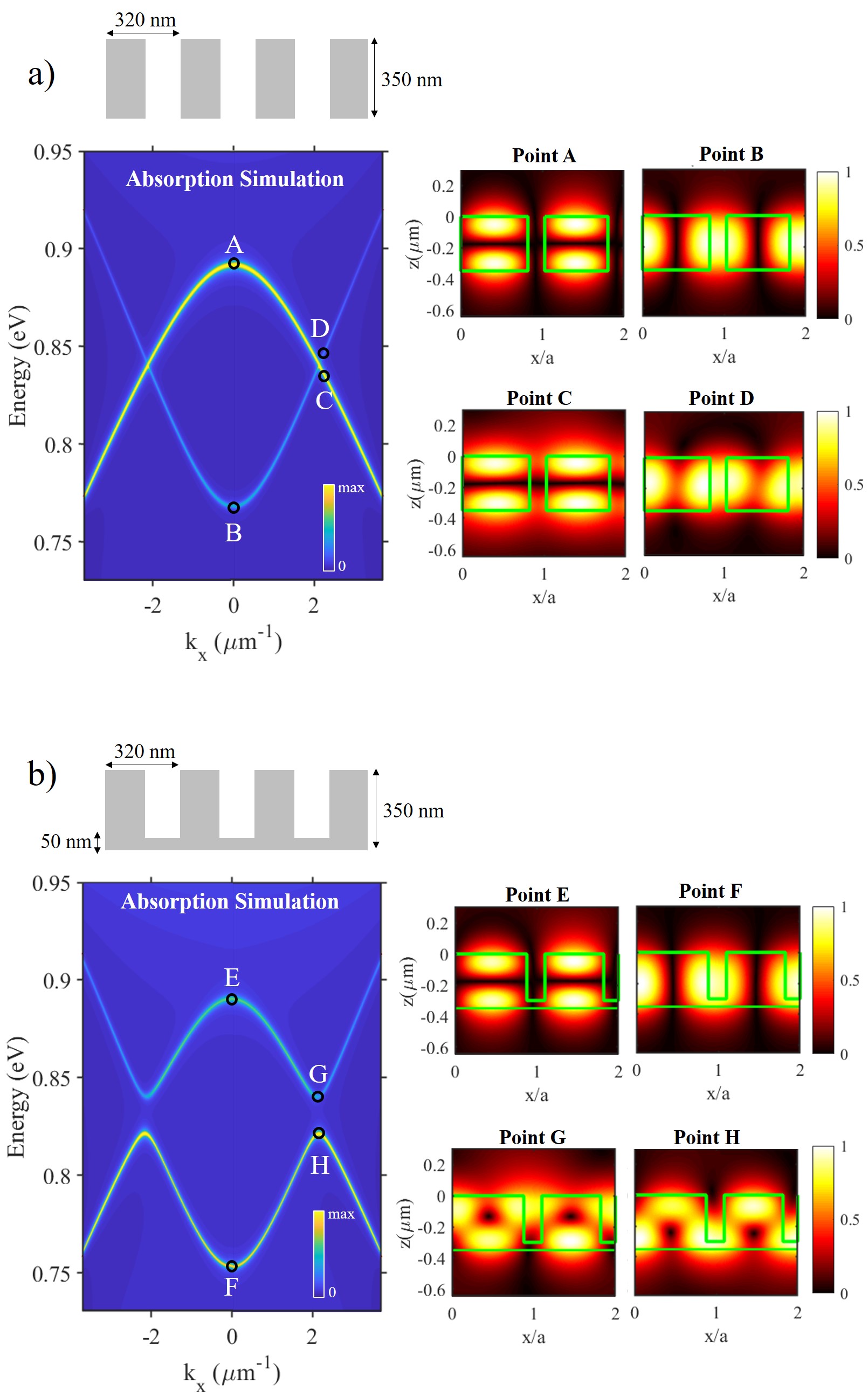}
\end{center}
\caption{Energy-momentum dispersion of two HCGs whose sole differences lie in the etch-depth: a) a fully-etched 350nm-thick Silicon HCG, b) a partially etched HCG (etch-depth: 300nm). We clearly observe the mode mixing between two bands originally with different parities ($\lambda$=1.5~$\mu$m, $k$=± 2~$\mu$m$^{-1}$). Breaking the vertical symmetry enables hybridizing these two previously orthogonal modes. Simulations were obtained with rigorous coupled-wave analysis (DiffractMODE, RSoft) }
\label{fig1}
\end{figure}

In Fig.~1 and in the rest of the paper (unless mentioned otherwise), as we are mainly interested in radiative modes, we introduce a double-period perturbation in the designs to fold the modes from the edge (X point) to the center ($\Gamma$ point) of the first Brillouin zone~\cite{combrie2010vertical}. This perturbation turns waveguided modes (below the light line) into radiative modes (above the light line), thus making them accessible via free-space illumination and collection. The perturbation magnitude is set by a factor $\alpha$ that governs the radiative lifetime and hence the quality factor of the mode~\cite{nguyen2018symmetry}. Note that this double-period structuration of HCG was recently given the alternative name of "dimerized HCG"~\cite{overvig2018dimerized}, hence the devices presented here can be viewed as dimerized HCGs with vertical symmetry breaking.  

\subsection{Engineering dispersion and density of states}
In a high index membrane supporting two guided modes, breaking vertical symmetry through partial etching provides access to a wide variety of dispersion characteristics~\cite{nguyen2018symmetry}. Controlling the etch depth allows for adjusting the coupling not only between the propagating and contrapropagating modes of each guided mode but also, through vertical symmetry breaking, between the even fundamental and odd first-excited guided modes. A progressive decrease of the etch depth generates successively a Dirac cone dispersion, then ultra-flat band conditions where both the group velocity and the second derivative or curvature of the dispersion characteristics are null, followed by M- and W-shaped dispersions. 
\begin{figure}[ht]
\begin{center}
\includegraphics[width=8.5cm]{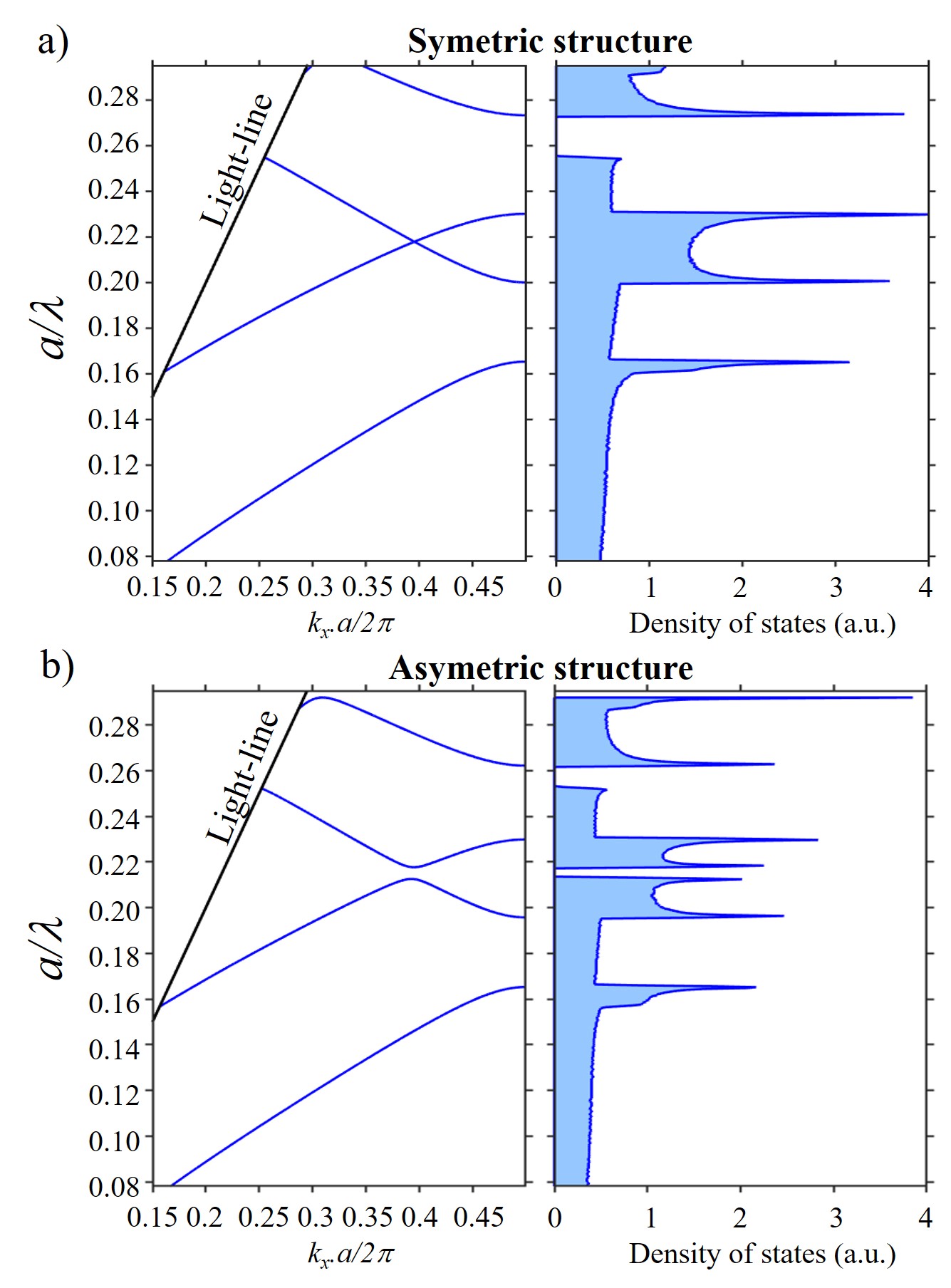}
\end{center}
\caption{Band diagrams and optical density of states calculated for a) vertically symmetric HCG and b) symmetry-broken structure.}
\label{fig2}
\end{figure}

Note that the LDOS is directly linked to the slopes and curvatures of the different optical modes. Therefore, tuning the dispersion's shape may enable tailoring the LDOS at an arbitrary point in energy-momentum space. To study this effect, the 1D LDOS ($k_x$ direction) have been computed for both structures shown in Fig.~1. 
The band structure has been obtained using a commercial Plane Wave Expansion software (Bandsolve, RSoft), and the LDOS have been deduced from the dispersion by summing the number of modes located at $w$ over a $dw$ window. As explained before, and shown in Fig.2, the broken vertical symmetry opens up a bandgap at the anti-crossing point, and because of the vanishing group velocity ($V_g$=$dw$/$dk$ tends to zero), the LDOS is sharply increased at the edges of the new hybridized modes. With this relatively simple method, one can therefore tailor the optical dispersion of a comb-like HCG to a wealth of different shapes. By simply modifying the etch depth, one can tune the slope and curvature of the dispersion at some specific $k$ values. This enables precise adjustment of the high LDOS region in energy and momentum space (i.e. choosing an appropriate set of $k_x$ and $\lambda$ values to enhance the LDOS).

In the following, we demonstrate how this concept can be used to enhance, at specific wavelength and direction, on-chip light emission from silicon at room-temperature.

\section{Results}

We specifically designed a HCG structure with broken vertical symmetry that contains different hybridized modes in the wavelength range where silicon emits light. The structure, sketched in Fig.~3a, is based on the following vertical stack: crystalline silicon on silica/insulator/amorphous-silicon. The insulator layer (Y$_2$O$_3$ here) is used as an etch-stop layer to ensure we precisely control the etch depth of the resulting device. The period ($a$) of the photonic membrane is 335~nm, the air filling factor ($FF$) is 0.375, and the double-period perturbation factor ($\alpha$) is 0.1. 
The theoretical absorption and field distributions of the structure were modeled using Rigorous Coupled-Wave Analysis (DiffractMODE, RSoft). As can be seen in the right part of Fig.~3c, the optical energy-momentum dispersion diagram contains five bands with a variety of shapes and curvatures; namely two parabolic bands, two M-shaped bands, and one flatband. The coupling between bands 3, 4, and 5 is responsible for the exotic shapes of these three bands. In the four upper bands, most of the field is confined in either the c-Si layer or the a-Si gratings (see Fig.~3d). 

\begin{figure}[t]
\begin{center}
\includegraphics[width=9cm]{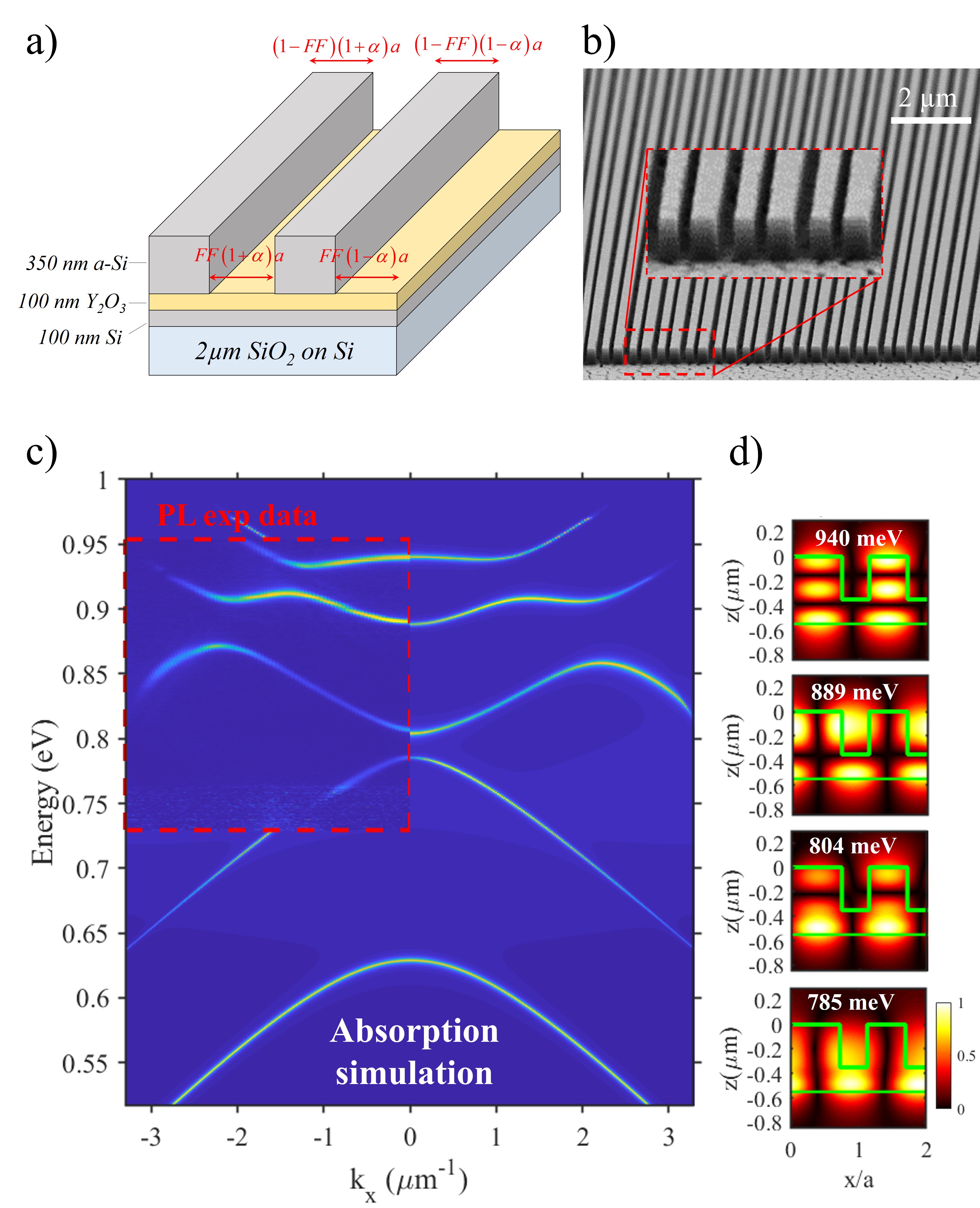}
\end{center}
\caption{a)  Sketch of the designed broken-symmetry HCG. b) Scanning electron microscope image of the fabricated device. c) Simulated absorption in the structure, together with the Energy-Momentum-resolved PL measurement shown as inset in the upper left part. d) Cross-sections of the devices showing the field distribution intensity ($E_y$) for the four different bands at $k$=0}
\label{fig3}
\end{figure}

Because of its indirect bandgap, silicon is a very poor light emitter at room-temperature. However, it is known that defects in both crystalline and amorphous silicon produce weak light emission below the bandgap around 1.3 $\mu$m~\cite{lo2011room,fujita2008light}. We leverage these weak emitters to probe the optical modes within the symmetry-broken HCG and to study how their emission is affected by the engineered optical dispersion of the structure.

\subsection{Fabrication}

We started from a commercial SOI substrate with a 2 $\mu$m oxide and a 100 nm top silicon layer. Then, a 100-nm thick Y$_2$O$_3$ layer was deposited by e-beam evaporation. It was then annealed under a flux of O$_2$ (0.5 lpm) for 1 hour at 900$^{o}$C to ensure the stoichiometry and crystallinity of the Y$_2$O$_3$. This layer was used as an etch stop to precisely control the etch depth of the gratings. We subsequently deposited 352 nm of amorphous silicon by plasma-enhanced chemical vapor deposition (PECVD) using an SiH$_4$ precursor and helium as the plasma gas. The plasma was set by an RF signal at 25W, the pressure in the chamber was kept at 2 Torr, and the substrate temperature was set to 300$^{o}$C. 
A 100~nm-thick hydrogen silsesquioxane (HSQ) resist was then spun on the sample and baked at 80$^{o}$C for 4 minutes. The resist was then exposed by electron-beam lithography, and the patterns were transferred to the a-Si by inductively-coupled reactive ion etching (ICP-RIE) using a mixture of Cl$_2$ and Ar.

A conventional inverted microscope (Nikon Eclipse Ti) was used to measure the photoluminescence of the samples. An oil objective (Nikon Plan Fluor, 100x, 1.3 NA) was used to focus the pump light and collect the emitted light. A diode-pumped, frequency-doubled Nd:YVO$_4$ laser at 532-nm (Verdi Coherent) was used to optically excite the emitters. The resulting NIR emission was then dispersed by an aberration-corrected Schmidt-Czerny-Turner spectrometer (Princeton Instruments, IsoPlane SCT-320), and subsequently measured by a high-resolution 2D InGaAs focal plane array (Princeton Instruments, PIoNIR 640).

For energy-momentum spectroscopy measurements, a Bertrand lens of focal length 250-mm was placed between the objective and tube lens to image the objective's back focal plane onto the entrance slit of the spectrograph. As suggested in~\cite{kurvits2015comparative}, this configuration simplifies the optical alignment process and significantly improves Fourier imaging quality. A polarizer between the Bertrand lens and objective allowed the polarization of emitted light to be filtered. In all the following results, the polarizer was placed so that only TE-polarized light emission is measured. The gratings were aligned perpendicular to the slits of the spectrograph, hence allowing to image the dispersion along k$_x$ while k$_y$=0.   

\subsection{Increasing the LDOS}

The left side inset in Fig.~3c shows the experimental photoluminescence (PL) measured using the energy-momentum spectroscopy setup. All measurements were done at room temperature. We find a very good agreement between the measured PL and the simulated optical absorption shown on the right-side of Fig.~3c. As expected, the silicon PL is coupled to the available guided modes and subsequently redirected to free-space inside the light cone owing to the double period corrugation. No significant PL emission is detected outside of the predicted bands. Note that, to obtain a perfect agreement between measurement and simulations, the geometric parameters had to be slightly adjusted in the model (period $a$=329.5 instead of 335nm, $FF$=0.355 instead of 0.375, and $\alpha$=0.125 instead of 0.1). Such a correction is due to a small mismatch between the dimensions of the designed and the fabricated structures.

\begin{figure}[!ht]
\begin{center}
\includegraphics[width=8.5cm]{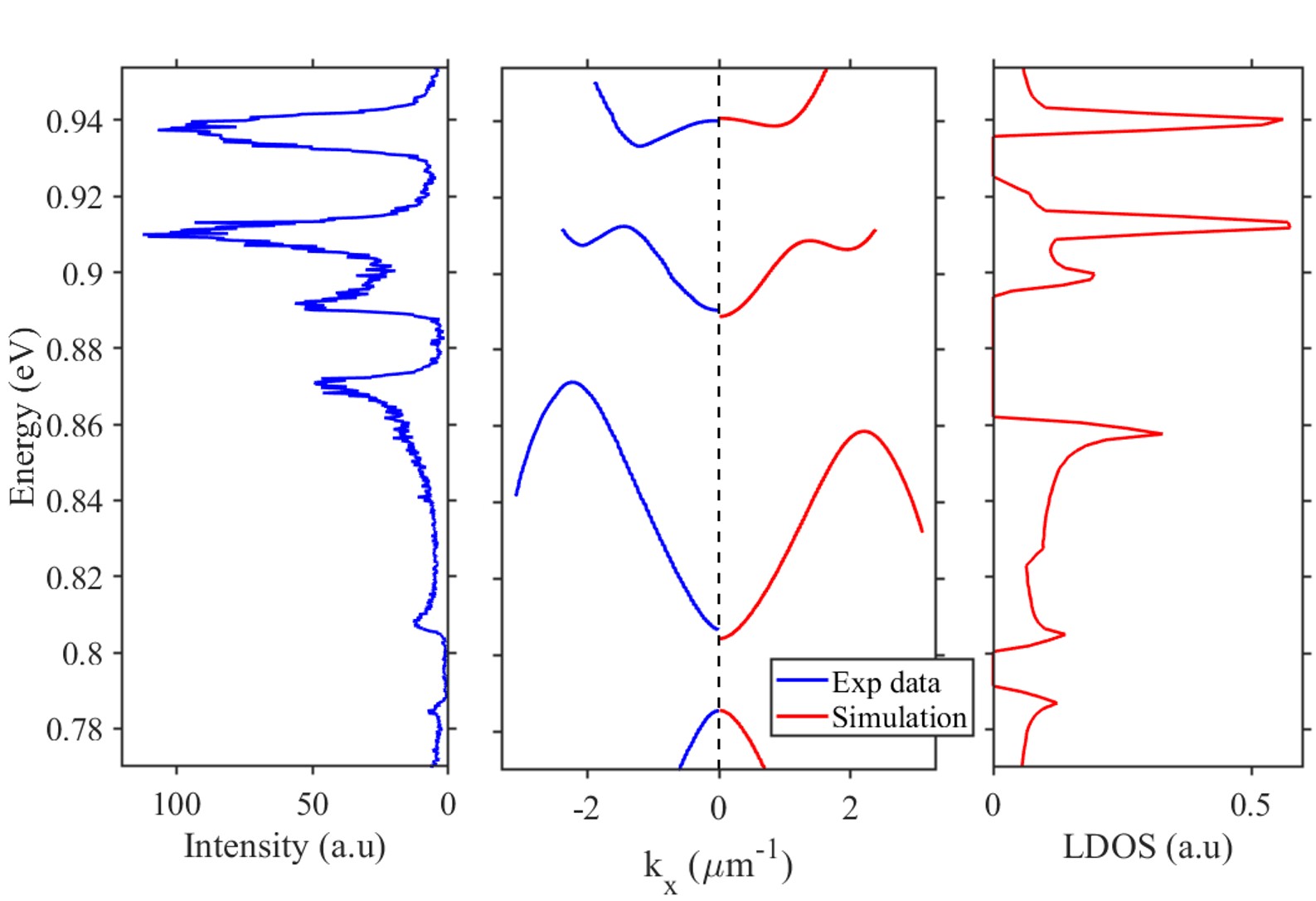}
\end{center}
\caption{Experimental and simulated energy-momentum-resolved PL.  Left and right: measured 1D LDOS and calculated 1D LDOS.}
\label{fig4}
\end{figure}

By summing all the angular contributions to the measured PL, or in other words by adding the detected intensities for all $k_x$-values as a function of energy, it is possible to reconstruct a spectrum that is an experimental measurement of the 1D LDOS. Fig.~4 shows the energy-momentum-resolved PL measurement together with the experimental 1D LDOS (left) and the calculated 1D LDOS (right).

A good qualitative agreement is observed between the measured and calculated LDOS. For one dimensional photonic crystals, the 1D LDOS is inversely proportional to the group velocity $V_g$ of the dispersion, i.e. the slope $dw$/$dk$ of the corresponding eigenmode.
This is indeed verified here, as high LDOS are observed at energies where the group velocity vanishes, i.e. at the extrema of the photonic bands.   
Thus, following these simple design principles, one could readily engineer modes with high LDOS at a desired wavelength.

In this particular case of 1D structuration, the total DOS is not expected to significantly change as compared to an unpatterned sample. However, as we filter out the $k_y$ contribution by measuring the dispersion around $k_y$=0, we are able to provide a good picture of the LDOS variation along the $k_x$ direction. The same approach was used to calculate the theoretical LDOS. Note that, in the present paper, we focused for simplicity in the 1D case, but it is straightforward to obtain a “global” modification of DOS by patterning the sample in 2D (e.g., using pillars array or air holes photonic crystals) to tailor the dispersion along both k$_x$ and k$_y$.

\subsection{Tailoring the direction of light emission}

Displaying the experimental PL intensity in a 3D-plot as a function of the energy and $k_x$, we can clearly visualize the angular distribution of light emission for the different photonic bands (see Fig.~5). As can be seen in Fig.~5b, the different bands have distinct intensity variations as a function of angle, each set by the shape of the optical dispersion. The flat-band produces an emission whose intensity is homogeneously distributed between $k$ = 0 to $k$ = 1 $\mu$m$^{-1}$, whereas the “M-band” beams the majority of its emission at high angles ($k_x$ $\ge$ 2 $\mu$m$^{-1}$). On the contrary, the parabolic band directs all its emission at normal incidence. These different bands clearly exemplify the variety of angular distributions and LDOS that can be obtained with such hybridized modes.
Furthermore, by simply changing geometric parameters such as the period or the filling factor, one could tune the angular distribution of these modes.

\begin{figure}[!ht]
\begin{center}
\includegraphics[width=8.5cm]{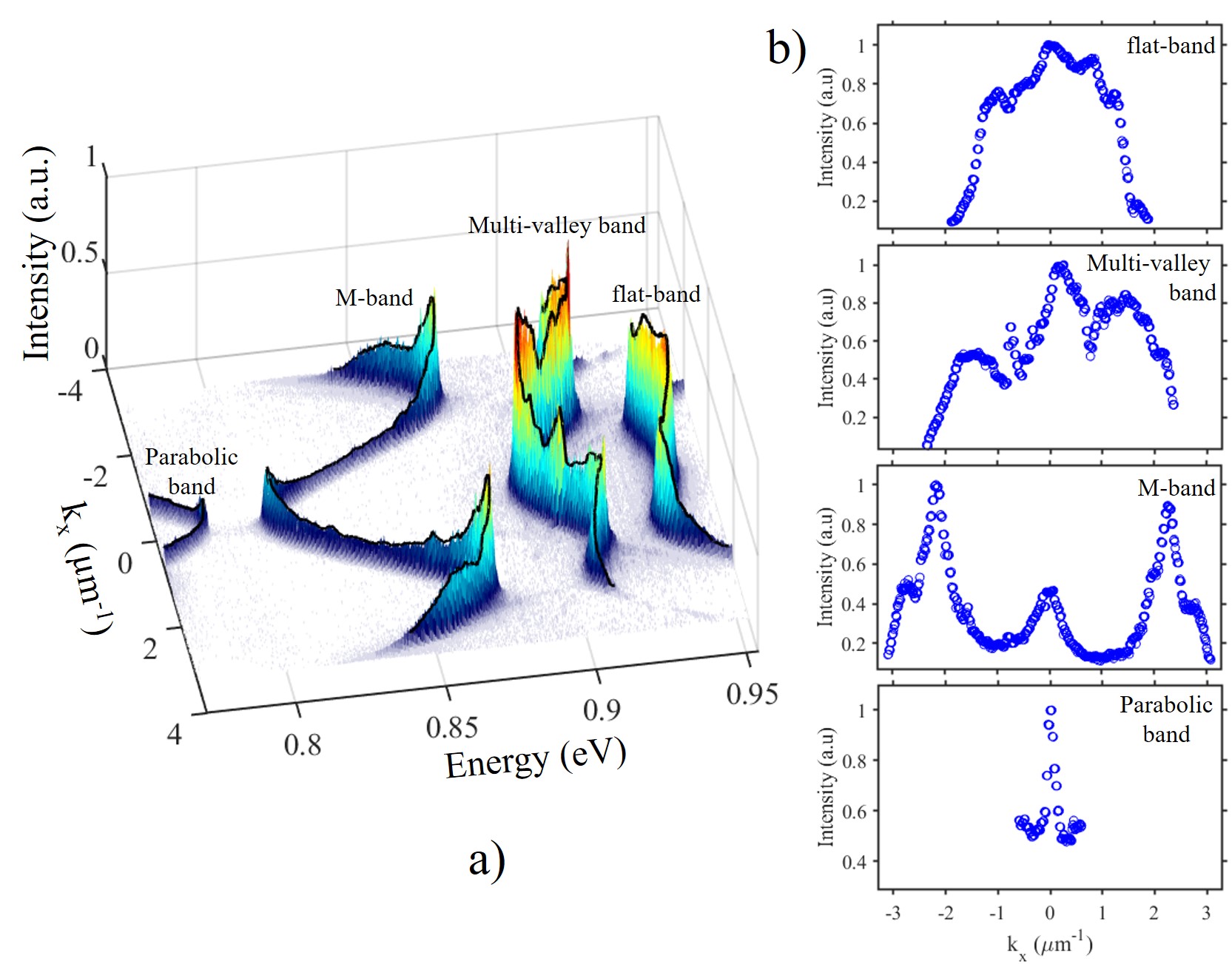}
\end{center}
\caption{a) Experimental energy-momentum-resolved PL intensity of the vertical symmetry-broken HCG. b) Normalized PL intensity as a function of $k_x$ for each band of the symmetry-broken HCG.}
\label{fig5}
\end{figure}

\section{Conclusion}
We have shown that vertical symmetry-breaking in HCG enables mixing optical modes of different parity. These new hybridized modes show exotic energy-momentum dispersion with band shapes that can be tuned by adjusting geometrical parameters. This controllable modifications of optical dispersion is particularly useful for simultaneously tailoring the LDOS and angle of light emission at specific wavelengths. Light emission can therefore be preferentially funneled into specifically chosen areas of the energy-momentum domain, at the expense of unwanted loss-channels.

Among numerous applications, such a platform could be readily used as an open cavity for efficiently extracting and redirecting weak or single-photon emitters on-chip such as nitrogen-vacancy centers in diamond~\cite{kurtsiefer2000stable} or G-centers in silicon~\cite{beaufils2018optical} into useful optical modes (radiative with an angle, guided, etc..). On a more fundamental level, flatbands and multivalley dispersion (M- and W-shaped bands) are increasingly investigated for exotic photonic effects~\cite{sun2017multivalley,leykam2018perspective}. 

We also envision future tunable devices leveraging the presented concept. Indeed, a small change in filling-factor could easily redirect light emission to different modes, controllably modifying the angular emission pattern. For example, by using liquid crystals~\cite{sautter2015active}, piezoelectric materials~\cite{midolo2018nano} or phase-change materials~\cite{cueff2015dynamic,wang2016optically,rios2015integrated}, this approach could enable dynamically tuning the angle of emission for beam-steering applications. On the other hand, a homogeneously distributed light emission into a specific solid angle (such as in the flat-band case) could be attractive for domains of applications such as solid-state lighting and Visible Light Communication (VLC)~\cite{Koonen2018}.

\section*{Acknowledgment}

The authors would like to thank the staff from the NanoLyon Technical Platform for helping and supporting in all nanofabrication processes. This work is partly supported by the French National Research Agency (ANR) under the projects PICSEL, SNAPSHOT and POPEYE as well as the United States National Science Foundation (ECCS-1408009).



%

\bibliography{Refs}

\begin{thebibliography}{36}%
\makeatletter
\providecommand \@ifxundefined [1]{%
 \@ifx{#1\undefined}
}%
\providecommand \@ifnum [1]{%
 \ifnum #1\expandafter \@firstoftwo
 \else \expandafter \@secondoftwo
 \fi
}%
\providecommand \@ifx [1]{%
 \ifx #1\expandafter \@firstoftwo
 \else \expandafter \@secondoftwo
 \fi
}%
\providecommand \natexlab [1]{#1}%
\providecommand \enquote  [1]{``#1''}%
\providecommand \bibnamefont  [1]{#1}%
\providecommand \bibfnamefont [1]{#1}%
\providecommand \citenamefont [1]{#1}%
\providecommand \href@noop [0]{\@secondoftwo}%
\providecommand \href [0]{\begingroup \@sanitize@url \@href}%
\providecommand \@href[1]{\@@startlink{#1}\@@href}%
\providecommand \@@href[1]{\endgroup#1\@@endlink}%
\providecommand \@sanitize@url [0]{\catcode `\\12\catcode `\$12\catcode
  `\&12\catcode `\#12\catcode `\^12\catcode `\_12\catcode `\%12\relax}%
\providecommand \@@startlink[1]{}%
\providecommand \@@endlink[0]{}%
\providecommand \url  [0]{\begingroup\@sanitize@url \@url }%
\providecommand \@url [1]{\endgroup\@href {#1}{\urlprefix }}%
\providecommand \urlprefix  [0]{URL }%
\providecommand \Eprint [0]{\href }%
\providecommand \doibase [0]{http://dx.doi.org/}%
\providecommand \selectlanguage [0]{\@gobble}%
\providecommand \bibinfo  [0]{\@secondoftwo}%
\providecommand \bibfield  [0]{\@secondoftwo}%
\providecommand \translation [1]{[#1]}%
\providecommand \BibitemOpen [0]{}%
\providecommand \bibitemStop [0]{}%
\providecommand \bibitemNoStop [0]{.\EOS\space}%
\providecommand \EOS [0]{\spacefactor3000\relax}%
\providecommand \BibitemShut  [1]{\csname bibitem#1\endcsname}%
\let\auto@bib@innerbib\@empty
\bibitem [{\citenamefont {Soref}(2006)}]{soref2006past}%
  \BibitemOpen
  \bibfield  {author} {\bibinfo {author} {\bibfnamefont {R.}~\bibnamefont
  {Soref}},\ }\href@noop {} {\bibfield  {journal} {\bibinfo  {journal} {IEEE
  Journal of selected topics in quantum electronics}\ }\textbf {\bibinfo
  {volume} {12}},\ \bibinfo {pages} {1678} (\bibinfo {year}
  {2006})}\BibitemShut {NoStop}%
\bibitem [{\citenamefont {Lipson}(2005)}]{lipson2005guiding}%
  \BibitemOpen
  \bibfield  {author} {\bibinfo {author} {\bibfnamefont {M.}~\bibnamefont
  {Lipson}},\ }\href@noop {} {\bibfield  {journal} {\bibinfo  {journal}
  {Journal of Lightwave Technology}\ }\textbf {\bibinfo {volume} {23}},\
  \bibinfo {pages} {4222} (\bibinfo {year} {2005})}\BibitemShut {NoStop}%
\bibitem [{\citenamefont {Purcell}(1995)}]{purcell}%
  \BibitemOpen
  \bibfield  {author} {\bibinfo {author} {\bibfnamefont {E.~M.}\ \bibnamefont
  {Purcell}},\ }\bibfield  {booktitle} {\emph {\bibinfo {booktitle} {Confined
  Electrons and Photons}},\ }\href@noop {} {\ ,\ \bibinfo {pages} {839}
  (\bibinfo {year} {1995})}\BibitemShut {NoStop}%
\bibitem [{\citenamefont {Pelton}(2015)}]{pelton2015}%
  \BibitemOpen
  \bibfield  {author} {\bibinfo {author} {\bibfnamefont {M.}~\bibnamefont
  {Pelton}},\ }\href@noop {} {\bibfield  {journal} {\bibinfo  {journal} {Nature
  Photonics}\ }\textbf {\bibinfo {volume} {9}},\ \bibinfo {pages} {427}
  (\bibinfo {year} {2015})}\BibitemShut {NoStop}%
\bibitem [{\citenamefont {Lo~Savio}\ \emph {et~al.}(2011)\citenamefont
  {Lo~Savio}, \citenamefont {Portalupi}, \citenamefont {Gerace}, \citenamefont
  {Shakoor}, \citenamefont {Krauss}, \citenamefont {O’Faolain}, \citenamefont
  {Andreani},\ and\ \citenamefont {Galli}}]{lo2011room}%
  \BibitemOpen
  \bibfield  {author} {\bibinfo {author} {\bibfnamefont {R.}~\bibnamefont
  {Lo~Savio}}, \bibinfo {author} {\bibfnamefont {S.}~\bibnamefont {Portalupi}},
  \bibinfo {author} {\bibfnamefont {D.}~\bibnamefont {Gerace}}, \bibinfo
  {author} {\bibfnamefont {A.}~\bibnamefont {Shakoor}}, \bibinfo {author}
  {\bibfnamefont {T.}~\bibnamefont {Krauss}}, \bibinfo {author} {\bibfnamefont
  {L.}~\bibnamefont {O’Faolain}}, \bibinfo {author} {\bibfnamefont
  {L.}~\bibnamefont {Andreani}}, \ and\ \bibinfo {author} {\bibfnamefont
  {M.}~\bibnamefont {Galli}},\ }\href@noop {} {\bibfield  {journal} {\bibinfo
  {journal} {Applied Physics Letters}\ }\textbf {\bibinfo {volume} {98}},\
  \bibinfo {pages} {201106} (\bibinfo {year} {2011})}\BibitemShut {NoStop}%
\bibitem [{\citenamefont {Zelsmann}\ \emph {et~al.}(2003)\citenamefont
  {Zelsmann}, \citenamefont {Picard}, \citenamefont {Charvolin}, \citenamefont
  {Hadji}, \citenamefont {Heitzmann}, \citenamefont {Dal’Zotto},
  \citenamefont {Nier}, \citenamefont {Seassal}, \citenamefont {Rojo-Romeo},\
  and\ \citenamefont {Letartre}}]{zelsmann2003seventy}%
  \BibitemOpen
  \bibfield  {author} {\bibinfo {author} {\bibfnamefont {M.}~\bibnamefont
  {Zelsmann}}, \bibinfo {author} {\bibfnamefont {E.}~\bibnamefont {Picard}},
  \bibinfo {author} {\bibfnamefont {T.}~\bibnamefont {Charvolin}}, \bibinfo
  {author} {\bibfnamefont {E.}~\bibnamefont {Hadji}}, \bibinfo {author}
  {\bibfnamefont {M.}~\bibnamefont {Heitzmann}}, \bibinfo {author}
  {\bibfnamefont {B.}~\bibnamefont {Dal’Zotto}}, \bibinfo {author}
  {\bibfnamefont {M.}~\bibnamefont {Nier}}, \bibinfo {author} {\bibfnamefont
  {C.}~\bibnamefont {Seassal}}, \bibinfo {author} {\bibfnamefont
  {P.}~\bibnamefont {Rojo-Romeo}}, \ and\ \bibinfo {author} {\bibfnamefont
  {X.}~\bibnamefont {Letartre}},\ }\href@noop {} {\bibfield  {journal}
  {\bibinfo  {journal} {Applied physics letters}\ }\textbf {\bibinfo {volume}
  {83}},\ \bibinfo {pages} {2542} (\bibinfo {year} {2003})}\BibitemShut
  {NoStop}%
\bibitem [{\citenamefont {Fujita}\ \emph {et~al.}(2008)\citenamefont {Fujita},
  \citenamefont {Tanaka},\ and\ \citenamefont {Noda}}]{fujita2008light}%
  \BibitemOpen
  \bibfield  {author} {\bibinfo {author} {\bibfnamefont {M.}~\bibnamefont
  {Fujita}}, \bibinfo {author} {\bibfnamefont {Y.}~\bibnamefont {Tanaka}}, \
  and\ \bibinfo {author} {\bibfnamefont {S.}~\bibnamefont {Noda}},\ }\href@noop
  {} {\bibfield  {journal} {\bibinfo  {journal} {IEEE Journal of Selected
  Topics in Quantum Electronics}\ }\textbf {\bibinfo {volume} {14}},\ \bibinfo
  {pages} {1090} (\bibinfo {year} {2008})}\BibitemShut {NoStop}%
\bibitem [{\citenamefont {Lodahl}\ \emph {et~al.}(2004)\citenamefont {Lodahl},
  \citenamefont {Van~Driel}, \citenamefont {Nikolaev}, \citenamefont {Irman},
  \citenamefont {Overgaag}, \citenamefont {Vanmaekelbergh},\ and\ \citenamefont
  {Vos}}]{lodahl2004controlling}%
  \BibitemOpen
  \bibfield  {author} {\bibinfo {author} {\bibfnamefont {P.}~\bibnamefont
  {Lodahl}}, \bibinfo {author} {\bibfnamefont {A.~F.}\ \bibnamefont
  {Van~Driel}}, \bibinfo {author} {\bibfnamefont {I.~S.}\ \bibnamefont
  {Nikolaev}}, \bibinfo {author} {\bibfnamefont {A.}~\bibnamefont {Irman}},
  \bibinfo {author} {\bibfnamefont {K.}~\bibnamefont {Overgaag}}, \bibinfo
  {author} {\bibfnamefont {D.}~\bibnamefont {Vanmaekelbergh}}, \ and\ \bibinfo
  {author} {\bibfnamefont {W.~L.}\ \bibnamefont {Vos}},\ }\href@noop {}
  {\bibfield  {journal} {\bibinfo  {journal} {Nature}\ }\textbf {\bibinfo
  {volume} {430}},\ \bibinfo {pages} {654} (\bibinfo {year}
  {2004})}\BibitemShut {NoStop}%
\bibitem [{\citenamefont {Englund}\ \emph {et~al.}(2005)\citenamefont
  {Englund}, \citenamefont {Fattal}, \citenamefont {Waks}, \citenamefont
  {Solomon}, \citenamefont {Zhang}, \citenamefont {Nakaoka}, \citenamefont
  {Arakawa}, \citenamefont {Yamamoto},\ and\ \citenamefont
  {Vu{\v{c}}kovi{\'c}}}]{englund2005controlling}%
  \BibitemOpen
  \bibfield  {author} {\bibinfo {author} {\bibfnamefont {D.}~\bibnamefont
  {Englund}}, \bibinfo {author} {\bibfnamefont {D.}~\bibnamefont {Fattal}},
  \bibinfo {author} {\bibfnamefont {E.}~\bibnamefont {Waks}}, \bibinfo {author}
  {\bibfnamefont {G.}~\bibnamefont {Solomon}}, \bibinfo {author} {\bibfnamefont
  {B.}~\bibnamefont {Zhang}}, \bibinfo {author} {\bibfnamefont
  {T.}~\bibnamefont {Nakaoka}}, \bibinfo {author} {\bibfnamefont
  {Y.}~\bibnamefont {Arakawa}}, \bibinfo {author} {\bibfnamefont
  {Y.}~\bibnamefont {Yamamoto}}, \ and\ \bibinfo {author} {\bibfnamefont
  {J.}~\bibnamefont {Vu{\v{c}}kovi{\'c}}},\ }\href@noop {} {\bibfield
  {journal} {\bibinfo  {journal} {Physical review letters}\ }\textbf {\bibinfo
  {volume} {95}},\ \bibinfo {pages} {013904} (\bibinfo {year}
  {2005})}\BibitemShut {NoStop}%
\bibitem [{\citenamefont {Fujita}\ \emph {et~al.}(2005)\citenamefont {Fujita},
  \citenamefont {Takahashi}, \citenamefont {Tanaka}, \citenamefont {Asano},\
  and\ \citenamefont {Noda}}]{fujita2005simultaneous}%
  \BibitemOpen
  \bibfield  {author} {\bibinfo {author} {\bibfnamefont {M.}~\bibnamefont
  {Fujita}}, \bibinfo {author} {\bibfnamefont {S.}~\bibnamefont {Takahashi}},
  \bibinfo {author} {\bibfnamefont {Y.}~\bibnamefont {Tanaka}}, \bibinfo
  {author} {\bibfnamefont {T.}~\bibnamefont {Asano}}, \ and\ \bibinfo {author}
  {\bibfnamefont {S.}~\bibnamefont {Noda}},\ }\href@noop {} {\bibfield
  {journal} {\bibinfo  {journal} {Science}\ }\textbf {\bibinfo {volume}
  {308}},\ \bibinfo {pages} {1296} (\bibinfo {year} {2005})}\BibitemShut
  {NoStop}%
\bibitem [{\citenamefont {Noda}\ \emph {et~al.}(2007)\citenamefont {Noda},
  \citenamefont {Fujita},\ and\ \citenamefont {Asano}}]{noda2007spontaneous}%
  \BibitemOpen
  \bibfield  {author} {\bibinfo {author} {\bibfnamefont {S.}~\bibnamefont
  {Noda}}, \bibinfo {author} {\bibfnamefont {M.}~\bibnamefont {Fujita}}, \ and\
  \bibinfo {author} {\bibfnamefont {T.}~\bibnamefont {Asano}},\ }\href@noop {}
  {\bibfield  {journal} {\bibinfo  {journal} {Nature photonics}\ }\textbf
  {\bibinfo {volume} {1}},\ \bibinfo {pages} {449} (\bibinfo {year}
  {2007})}\BibitemShut {NoStop}%
\bibitem [{\citenamefont {Makarova}\ \emph {et~al.}(2010)\citenamefont
  {Makarova}, \citenamefont {Gong}, \citenamefont {Cheng}, \citenamefont
  {Nishi}, \citenamefont {Yerci}, \citenamefont {Li}, \citenamefont
  {Dal~Negro},\ and\ \citenamefont {Vuckovic}}]{makarova2010photonic}%
  \BibitemOpen
  \bibfield  {author} {\bibinfo {author} {\bibfnamefont {M.}~\bibnamefont
  {Makarova}}, \bibinfo {author} {\bibfnamefont {Y.}~\bibnamefont {Gong}},
  \bibinfo {author} {\bibfnamefont {S.-L.}\ \bibnamefont {Cheng}}, \bibinfo
  {author} {\bibfnamefont {Y.}~\bibnamefont {Nishi}}, \bibinfo {author}
  {\bibfnamefont {S.}~\bibnamefont {Yerci}}, \bibinfo {author} {\bibfnamefont
  {R.}~\bibnamefont {Li}}, \bibinfo {author} {\bibfnamefont {L.}~\bibnamefont
  {Dal~Negro}}, \ and\ \bibinfo {author} {\bibfnamefont {J.}~\bibnamefont
  {Vuckovic}},\ }\href@noop {} {\bibfield  {journal} {\bibinfo  {journal} {IEEE
  Journal of Selected Topics in Quantum Electronics}\ }\textbf {\bibinfo
  {volume} {16}},\ \bibinfo {pages} {132} (\bibinfo {year} {2010})}\BibitemShut
  {NoStop}%
\bibitem [{\citenamefont {Tsakmakidis}\ \emph {et~al.}(2016)\citenamefont
  {Tsakmakidis}, \citenamefont {Boyd}, \citenamefont {Yablonovitch},\ and\
  \citenamefont {Zhang}}]{tsakmakidis2016large}%
  \BibitemOpen
  \bibfield  {author} {\bibinfo {author} {\bibfnamefont {K.~L.}\ \bibnamefont
  {Tsakmakidis}}, \bibinfo {author} {\bibfnamefont {R.~W.}\ \bibnamefont
  {Boyd}}, \bibinfo {author} {\bibfnamefont {E.}~\bibnamefont {Yablonovitch}},
  \ and\ \bibinfo {author} {\bibfnamefont {X.}~\bibnamefont {Zhang}},\
  }\href@noop {} {\bibfield  {journal} {\bibinfo  {journal} {Optics express}\
  }\textbf {\bibinfo {volume} {24}},\ \bibinfo {pages} {17916} (\bibinfo {year}
  {2016})}\BibitemShut {NoStop}%
\bibitem [{\citenamefont {Taminiau}\ \emph {et~al.}(2008)\citenamefont
  {Taminiau}, \citenamefont {Stefani}, \citenamefont {Segerink},\ and\
  \citenamefont {Van~Hulst}}]{taminiau2008optical}%
  \BibitemOpen
  \bibfield  {author} {\bibinfo {author} {\bibfnamefont {T.}~\bibnamefont
  {Taminiau}}, \bibinfo {author} {\bibfnamefont {F.}~\bibnamefont {Stefani}},
  \bibinfo {author} {\bibfnamefont {F.~B.}\ \bibnamefont {Segerink}}, \ and\
  \bibinfo {author} {\bibfnamefont {N.}~\bibnamefont {Van~Hulst}},\ }\href@noop
  {} {\bibfield  {journal} {\bibinfo  {journal} {Nature Photonics}\ }\textbf
  {\bibinfo {volume} {2}},\ \bibinfo {pages} {234} (\bibinfo {year}
  {2008})}\BibitemShut {NoStop}%
\bibitem [{\citenamefont {Curto}\ \emph {et~al.}(2010)\citenamefont {Curto},
  \citenamefont {Volpe}, \citenamefont {Taminiau}, \citenamefont {Kreuzer},
  \citenamefont {Quidant},\ and\ \citenamefont {van
  Hulst}}]{curto2010unidirectional}%
  \BibitemOpen
  \bibfield  {author} {\bibinfo {author} {\bibfnamefont {A.~G.}\ \bibnamefont
  {Curto}}, \bibinfo {author} {\bibfnamefont {G.}~\bibnamefont {Volpe}},
  \bibinfo {author} {\bibfnamefont {T.~H.}\ \bibnamefont {Taminiau}}, \bibinfo
  {author} {\bibfnamefont {M.~P.}\ \bibnamefont {Kreuzer}}, \bibinfo {author}
  {\bibfnamefont {R.}~\bibnamefont {Quidant}}, \ and\ \bibinfo {author}
  {\bibfnamefont {N.~F.}\ \bibnamefont {van Hulst}},\ }\href@noop {} {\bibfield
   {journal} {\bibinfo  {journal} {Science}\ }\textbf {\bibinfo {volume}
  {329}},\ \bibinfo {pages} {930} (\bibinfo {year} {2010})}\BibitemShut
  {NoStop}%
\bibitem [{\citenamefont {Lozano}\ \emph {et~al.}(2013)\citenamefont {Lozano},
  \citenamefont {Louwers}, \citenamefont {Rodr{\'\i}guez}, \citenamefont
  {Murai}, \citenamefont {Jansen}, \citenamefont {Verschuuren},\ and\
  \citenamefont {Rivas}}]{lozano2013plasmonics}%
  \BibitemOpen
  \bibfield  {author} {\bibinfo {author} {\bibfnamefont {G.}~\bibnamefont
  {Lozano}}, \bibinfo {author} {\bibfnamefont {D.~J.}\ \bibnamefont {Louwers}},
  \bibinfo {author} {\bibfnamefont {S.~R.}\ \bibnamefont {Rodr{\'\i}guez}},
  \bibinfo {author} {\bibfnamefont {S.}~\bibnamefont {Murai}}, \bibinfo
  {author} {\bibfnamefont {O.~T.}\ \bibnamefont {Jansen}}, \bibinfo {author}
  {\bibfnamefont {M.~A.}\ \bibnamefont {Verschuuren}}, \ and\ \bibinfo {author}
  {\bibfnamefont {J.~G.}\ \bibnamefont {Rivas}},\ }\href@noop {} {\bibfield
  {journal} {\bibinfo  {journal} {Light: Science \& Applications}\ }\textbf
  {\bibinfo {volume} {2}},\ \bibinfo {pages} {e66} (\bibinfo {year}
  {2013})}\BibitemShut {NoStop}%
\bibitem [{\citenamefont {Hancu}\ \emph {et~al.}(2013)\citenamefont {Hancu},
  \citenamefont {Curto}, \citenamefont {Castro-L{\'o}pez}, \citenamefont
  {Kuttge},\ and\ \citenamefont {van Hulst}}]{hancu2013multipolar}%
  \BibitemOpen
  \bibfield  {author} {\bibinfo {author} {\bibfnamefont {I.~M.}\ \bibnamefont
  {Hancu}}, \bibinfo {author} {\bibfnamefont {A.~G.}\ \bibnamefont {Curto}},
  \bibinfo {author} {\bibfnamefont {M.}~\bibnamefont {Castro-L{\'o}pez}},
  \bibinfo {author} {\bibfnamefont {M.}~\bibnamefont {Kuttge}}, \ and\ \bibinfo
  {author} {\bibfnamefont {N.~F.}\ \bibnamefont {van Hulst}},\ }\href@noop {}
  {\bibfield  {journal} {\bibinfo  {journal} {Nano letters}\ }\textbf {\bibinfo
  {volume} {14}},\ \bibinfo {pages} {166} (\bibinfo {year} {2013})}\BibitemShut
  {NoStop}%
\bibitem [{\citenamefont {Lee}\ \emph {et~al.}(2011)\citenamefont {Lee},
  \citenamefont {Chen}, \citenamefont {Eghlidi}, \citenamefont {Kukura},
  \citenamefont {Lettow}, \citenamefont {Renn}, \citenamefont {Sandoghdar},\
  and\ \citenamefont {G{\"o}tzinger}}]{lee2011planar}%
  \BibitemOpen
  \bibfield  {author} {\bibinfo {author} {\bibfnamefont {K.}~\bibnamefont
  {Lee}}, \bibinfo {author} {\bibfnamefont {X.}~\bibnamefont {Chen}}, \bibinfo
  {author} {\bibfnamefont {H.}~\bibnamefont {Eghlidi}}, \bibinfo {author}
  {\bibfnamefont {P.}~\bibnamefont {Kukura}}, \bibinfo {author} {\bibfnamefont
  {R.}~\bibnamefont {Lettow}}, \bibinfo {author} {\bibfnamefont
  {A.}~\bibnamefont {Renn}}, \bibinfo {author} {\bibfnamefont {V.}~\bibnamefont
  {Sandoghdar}}, \ and\ \bibinfo {author} {\bibfnamefont {S.}~\bibnamefont
  {G{\"o}tzinger}},\ }\href@noop {} {\bibfield  {journal} {\bibinfo  {journal}
  {Nature Photonics}\ }\textbf {\bibinfo {volume} {5}},\ \bibinfo {pages} {166}
  (\bibinfo {year} {2011})}\BibitemShut {NoStop}%
\bibitem [{\citenamefont {Cihan}\ \emph {et~al.}(2018)\citenamefont {Cihan},
  \citenamefont {Curto}, \citenamefont {Raza}, \citenamefont {Kik},\ and\
  \citenamefont {Brongersma}}]{cihan2018silicon}%
  \BibitemOpen
  \bibfield  {author} {\bibinfo {author} {\bibfnamefont {A.~F.}\ \bibnamefont
  {Cihan}}, \bibinfo {author} {\bibfnamefont {A.~G.}\ \bibnamefont {Curto}},
  \bibinfo {author} {\bibfnamefont {S.}~\bibnamefont {Raza}}, \bibinfo {author}
  {\bibfnamefont {P.~G.}\ \bibnamefont {Kik}}, \ and\ \bibinfo {author}
  {\bibfnamefont {M.~L.}\ \bibnamefont {Brongersma}},\ }\href@noop {}
  {\bibfield  {journal} {\bibinfo  {journal} {Nature Photonics}\ }\textbf
  {\bibinfo {volume} {12}},\ \bibinfo {pages} {284} (\bibinfo {year}
  {2018})}\BibitemShut {NoStop}%
\bibitem [{\citenamefont {Staude}\ \emph {et~al.}(2013)\citenamefont {Staude},
  \citenamefont {Miroshnichenko}, \citenamefont {Decker}, \citenamefont
  {Fofang}, \citenamefont {Liu}, \citenamefont {Gonzales}, \citenamefont
  {Dominguez}, \citenamefont {Luk}, \citenamefont {Neshev}, \citenamefont
  {Brener} \emph {et~al.}}]{staude2013tailoring}%
  \BibitemOpen
  \bibfield  {author} {\bibinfo {author} {\bibfnamefont {I.}~\bibnamefont
  {Staude}}, \bibinfo {author} {\bibfnamefont {A.~E.}\ \bibnamefont
  {Miroshnichenko}}, \bibinfo {author} {\bibfnamefont {M.}~\bibnamefont
  {Decker}}, \bibinfo {author} {\bibfnamefont {N.~T.}\ \bibnamefont {Fofang}},
  \bibinfo {author} {\bibfnamefont {S.}~\bibnamefont {Liu}}, \bibinfo {author}
  {\bibfnamefont {E.}~\bibnamefont {Gonzales}}, \bibinfo {author}
  {\bibfnamefont {J.}~\bibnamefont {Dominguez}}, \bibinfo {author}
  {\bibfnamefont {T.~S.}\ \bibnamefont {Luk}}, \bibinfo {author} {\bibfnamefont
  {D.~N.}\ \bibnamefont {Neshev}}, \bibinfo {author} {\bibfnamefont
  {I.}~\bibnamefont {Brener}},  \emph {et~al.},\ }\href@noop {} {\bibfield
  {journal} {\bibinfo  {journal} {ACS nano}\ }\textbf {\bibinfo {volume} {7}},\
  \bibinfo {pages} {7824} (\bibinfo {year} {2013})}\BibitemShut {NoStop}%
\bibitem [{\citenamefont {Ha}\ \emph {et~al.}(2018)\citenamefont {Ha},
  \citenamefont {Fu}, \citenamefont {Emani}, \citenamefont {Pan}, \citenamefont
  {Bakker}, \citenamefont {Paniagua-Dominguez},\ and\ \citenamefont
  {Kuznetsov}}]{ha2018lasing}%
  \BibitemOpen
  \bibfield  {author} {\bibinfo {author} {\bibfnamefont {S.~T.}\ \bibnamefont
  {Ha}}, \bibinfo {author} {\bibfnamefont {Y.~H.}\ \bibnamefont {Fu}}, \bibinfo
  {author} {\bibfnamefont {N.~K.}\ \bibnamefont {Emani}}, \bibinfo {author}
  {\bibfnamefont {Z.}~\bibnamefont {Pan}}, \bibinfo {author} {\bibfnamefont
  {R.~M.}\ \bibnamefont {Bakker}}, \bibinfo {author} {\bibfnamefont
  {R.}~\bibnamefont {Paniagua-Dominguez}}, \ and\ \bibinfo {author}
  {\bibfnamefont {A.~I.}\ \bibnamefont {Kuznetsov}},\ }\href@noop {} {\bibfield
   {journal} {\bibinfo  {journal} {arXiv preprint arXiv:1803.09993}\ }
  (\bibinfo {year} {2018})}\BibitemShut {NoStop}%
\bibitem [{\citenamefont {Chang-Hasnain}\ and\ \citenamefont
  {Yang}(2012)}]{chang2012high}%
  \BibitemOpen
  \bibfield  {author} {\bibinfo {author} {\bibfnamefont {C.~J.}\ \bibnamefont
  {Chang-Hasnain}}\ and\ \bibinfo {author} {\bibfnamefont {W.}~\bibnamefont
  {Yang}},\ }\href@noop {} {\bibfield  {journal} {\bibinfo  {journal} {Advances
  in Optics and Photonics}\ }\textbf {\bibinfo {volume} {4}},\ \bibinfo {pages}
  {379} (\bibinfo {year} {2012})}\BibitemShut {NoStop}%
\bibitem [{\citenamefont {Nguyen}\ \emph {et~al.}(2018)\citenamefont {Nguyen},
  \citenamefont {Dubois}, \citenamefont {Deschamps}, \citenamefont {Cueff},
  \citenamefont {Pardon}, \citenamefont {Leclercq}, \citenamefont {Seassal},
  \citenamefont {Letartre},\ and\ \citenamefont
  {Viktorovitch}}]{nguyen2018symmetry}%
  \BibitemOpen
  \bibfield  {author} {\bibinfo {author} {\bibfnamefont {H.~S.}\ \bibnamefont
  {Nguyen}}, \bibinfo {author} {\bibfnamefont {F.}~\bibnamefont {Dubois}},
  \bibinfo {author} {\bibfnamefont {T.}~\bibnamefont {Deschamps}}, \bibinfo
  {author} {\bibfnamefont {S.}~\bibnamefont {Cueff}}, \bibinfo {author}
  {\bibfnamefont {A.}~\bibnamefont {Pardon}}, \bibinfo {author} {\bibfnamefont
  {J.-L.}\ \bibnamefont {Leclercq}}, \bibinfo {author} {\bibfnamefont
  {C.}~\bibnamefont {Seassal}}, \bibinfo {author} {\bibfnamefont
  {X.}~\bibnamefont {Letartre}}, \ and\ \bibinfo {author} {\bibfnamefont
  {P.}~\bibnamefont {Viktorovitch}},\ }\href@noop {} {\bibfield  {journal}
  {\bibinfo  {journal} {Physical review letters}\ }\textbf {\bibinfo {volume}
  {120}},\ \bibinfo {pages} {066102} (\bibinfo {year} {2018})}\BibitemShut
  {NoStop}%
\bibitem [{\citenamefont {Combri{\'e}}\ \emph {et~al.}(2010)\citenamefont
  {Combri{\'e}}, \citenamefont {Colman}, \citenamefont {De~Rossi},
  \citenamefont {Mei} \emph {et~al.}}]{combrie2010vertical}%
  \BibitemOpen
  \bibfield  {author} {\bibinfo {author} {\bibfnamefont {S.}~\bibnamefont
  {Combri{\'e}}}, \bibinfo {author} {\bibfnamefont {P.}~\bibnamefont {Colman}},
  \bibinfo {author} {\bibfnamefont {A.}~\bibnamefont {De~Rossi}}, \bibinfo
  {author} {\bibfnamefont {T.}~\bibnamefont {Mei}},  \emph {et~al.},\
  }\href@noop {} {\bibfield  {journal} {\bibinfo  {journal} {Physical Review
  B}\ }\textbf {\bibinfo {volume} {82}},\ \bibinfo {pages} {075120} (\bibinfo
  {year} {2010})}\BibitemShut {NoStop}%
\bibitem [{\citenamefont {Overvig}\ \emph {et~al.}(2018)\citenamefont
  {Overvig}, \citenamefont {Shrestha},\ and\ \citenamefont
  {Yu}}]{overvig2018dimerized}%
  \BibitemOpen
  \bibfield  {author} {\bibinfo {author} {\bibfnamefont {A.~C.}\ \bibnamefont
  {Overvig}}, \bibinfo {author} {\bibfnamefont {S.}~\bibnamefont {Shrestha}}, \
  and\ \bibinfo {author} {\bibfnamefont {N.}~\bibnamefont {Yu}},\ }\href@noop
  {} {\bibfield  {journal} {\bibinfo  {journal} {Nanophotonics}\ }\textbf
  {\bibinfo {volume} {7}},\ \bibinfo {pages} {1157} (\bibinfo {year}
  {2018})}\BibitemShut {NoStop}%
\bibitem [{\citenamefont {Kurvits}\ \emph {et~al.}(2015)\citenamefont
  {Kurvits}, \citenamefont {Jiang},\ and\ \citenamefont
  {Zia}}]{kurvits2015comparative}%
  \BibitemOpen
  \bibfield  {author} {\bibinfo {author} {\bibfnamefont {J.~A.}\ \bibnamefont
  {Kurvits}}, \bibinfo {author} {\bibfnamefont {M.}~\bibnamefont {Jiang}}, \
  and\ \bibinfo {author} {\bibfnamefont {R.}~\bibnamefont {Zia}},\ }\href@noop
  {} {\bibfield  {journal} {\bibinfo  {journal} {JOSA A}\ }\textbf {\bibinfo
  {volume} {32}},\ \bibinfo {pages} {2082} (\bibinfo {year}
  {2015})}\BibitemShut {NoStop}%
\bibitem [{\citenamefont {Kurtsiefer}\ \emph {et~al.}(2000)\citenamefont
  {Kurtsiefer}, \citenamefont {Mayer}, \citenamefont {Zarda},\ and\
  \citenamefont {Weinfurter}}]{kurtsiefer2000stable}%
  \BibitemOpen
  \bibfield  {author} {\bibinfo {author} {\bibfnamefont {C.}~\bibnamefont
  {Kurtsiefer}}, \bibinfo {author} {\bibfnamefont {S.}~\bibnamefont {Mayer}},
  \bibinfo {author} {\bibfnamefont {P.}~\bibnamefont {Zarda}}, \ and\ \bibinfo
  {author} {\bibfnamefont {H.}~\bibnamefont {Weinfurter}},\ }\href@noop {}
  {\bibfield  {journal} {\bibinfo  {journal} {Physical review letters}\
  }\textbf {\bibinfo {volume} {85}},\ \bibinfo {pages} {290} (\bibinfo {year}
  {2000})}\BibitemShut {NoStop}%
\bibitem [{\citenamefont {Beaufils}\ \emph {et~al.}(2018)\citenamefont
  {Beaufils}, \citenamefont {Redjem}, \citenamefont {Rousseau}, \citenamefont
  {Jacques}, \citenamefont {Kuznetsov}, \citenamefont {Raynaud}, \citenamefont
  {Voisin}, \citenamefont {Benali}, \citenamefont {Herzig}, \citenamefont
  {Pezzagna}, \citenamefont {Meijer}, \citenamefont {Abbarchi},\ and\
  \citenamefont {Cassabois}}]{beaufils2018optical}%
  \BibitemOpen
  \bibfield  {author} {\bibinfo {author} {\bibfnamefont {C.}~\bibnamefont
  {Beaufils}}, \bibinfo {author} {\bibfnamefont {W.}~\bibnamefont {Redjem}},
  \bibinfo {author} {\bibfnamefont {E.}~\bibnamefont {Rousseau}}, \bibinfo
  {author} {\bibfnamefont {V.}~\bibnamefont {Jacques}}, \bibinfo {author}
  {\bibfnamefont {A.~Y.}\ \bibnamefont {Kuznetsov}}, \bibinfo {author}
  {\bibfnamefont {C.}~\bibnamefont {Raynaud}}, \bibinfo {author} {\bibfnamefont
  {C.}~\bibnamefont {Voisin}}, \bibinfo {author} {\bibfnamefont
  {A.}~\bibnamefont {Benali}}, \bibinfo {author} {\bibfnamefont
  {T.}~\bibnamefont {Herzig}}, \bibinfo {author} {\bibfnamefont
  {S.}~\bibnamefont {Pezzagna}}, \bibinfo {author} {\bibfnamefont
  {J.}~\bibnamefont {Meijer}}, \bibinfo {author} {\bibfnamefont
  {M.}~\bibnamefont {Abbarchi}}, \ and\ \bibinfo {author} {\bibfnamefont
  {G.}~\bibnamefont {Cassabois}},\ }\href@noop {} {\bibfield  {journal}
  {\bibinfo  {journal} {Physical Review B}\ }\textbf {\bibinfo {volume} {97}},\
  \bibinfo {pages} {035303} (\bibinfo {year} {2018})}\BibitemShut {NoStop}%
\bibitem [{\citenamefont {Sun}\ \emph {et~al.}(2017)\citenamefont {Sun},
  \citenamefont {Savenko}, \citenamefont {Flayac},\ and\ \citenamefont
  {Liew}}]{sun2017multivalley}%
  \BibitemOpen
  \bibfield  {author} {\bibinfo {author} {\bibfnamefont {M.}~\bibnamefont
  {Sun}}, \bibinfo {author} {\bibfnamefont {I.}~\bibnamefont {Savenko}},
  \bibinfo {author} {\bibfnamefont {H.}~\bibnamefont {Flayac}}, \ and\ \bibinfo
  {author} {\bibfnamefont {T.~C.~H.}\ \bibnamefont {Liew}},\ }\href@noop {}
  {\bibfield  {journal} {\bibinfo  {journal} {Scientific Reports}\ }\textbf
  {\bibinfo {volume} {7}},\ \bibinfo {pages} {45243} (\bibinfo {year}
  {2017})}\BibitemShut {NoStop}%
\bibitem [{\citenamefont {Leykam}\ and\ \citenamefont
  {Flach}(2018)}]{leykam2018perspective}%
  \BibitemOpen
  \bibfield  {author} {\bibinfo {author} {\bibfnamefont {D.}~\bibnamefont
  {Leykam}}\ and\ \bibinfo {author} {\bibfnamefont {S.}~\bibnamefont {Flach}},\
  }\href@noop {} {\bibfield  {journal} {\bibinfo  {journal} {APL Photonics}\
  }\textbf {\bibinfo {volume} {3}},\ \bibinfo {pages} {070901} (\bibinfo {year}
  {2018})}\BibitemShut {NoStop}%
\bibitem [{\citenamefont {Sautter}\ \emph {et~al.}(2015)\citenamefont
  {Sautter}, \citenamefont {Staude}, \citenamefont {Decker}, \citenamefont
  {Rusak}, \citenamefont {Neshev}, \citenamefont {Brener},\ and\ \citenamefont
  {Kivshar}}]{sautter2015active}%
  \BibitemOpen
  \bibfield  {author} {\bibinfo {author} {\bibfnamefont {J.}~\bibnamefont
  {Sautter}}, \bibinfo {author} {\bibfnamefont {I.}~\bibnamefont {Staude}},
  \bibinfo {author} {\bibfnamefont {M.}~\bibnamefont {Decker}}, \bibinfo
  {author} {\bibfnamefont {E.}~\bibnamefont {Rusak}}, \bibinfo {author}
  {\bibfnamefont {D.~N.}\ \bibnamefont {Neshev}}, \bibinfo {author}
  {\bibfnamefont {I.}~\bibnamefont {Brener}}, \ and\ \bibinfo {author}
  {\bibfnamefont {Y.~S.}\ \bibnamefont {Kivshar}},\ }\href@noop {} {\bibfield
  {journal} {\bibinfo  {journal} {ACS nano}\ }\textbf {\bibinfo {volume} {9}},\
  \bibinfo {pages} {4308} (\bibinfo {year} {2015})}\BibitemShut {NoStop}%
\bibitem [{\citenamefont {Midolo}\ \emph {et~al.}(2018)\citenamefont {Midolo},
  \citenamefont {Schliesser},\ and\ \citenamefont {Fiore}}]{midolo2018nano}%
  \BibitemOpen
  \bibfield  {author} {\bibinfo {author} {\bibfnamefont {L.}~\bibnamefont
  {Midolo}}, \bibinfo {author} {\bibfnamefont {A.}~\bibnamefont {Schliesser}},
  \ and\ \bibinfo {author} {\bibfnamefont {A.}~\bibnamefont {Fiore}},\
  }\href@noop {} {\bibfield  {journal} {\bibinfo  {journal} {Nature
  nanotechnology}\ }\textbf {\bibinfo {volume} {13}},\ \bibinfo {pages} {11}
  (\bibinfo {year} {2018})}\BibitemShut {NoStop}%
\bibitem [{\citenamefont {Cueff}\ \emph {et~al.}(2015)\citenamefont {Cueff},
  \citenamefont {Li}, \citenamefont {Zhou}, \citenamefont {Wong}, \citenamefont
  {Kurvits}, \citenamefont {Ramanathan},\ and\ \citenamefont
  {Zia}}]{cueff2015dynamic}%
  \BibitemOpen
  \bibfield  {author} {\bibinfo {author} {\bibfnamefont {S.}~\bibnamefont
  {Cueff}}, \bibinfo {author} {\bibfnamefont {D.}~\bibnamefont {Li}}, \bibinfo
  {author} {\bibfnamefont {Y.}~\bibnamefont {Zhou}}, \bibinfo {author}
  {\bibfnamefont {F.~J.}\ \bibnamefont {Wong}}, \bibinfo {author}
  {\bibfnamefont {J.~A.}\ \bibnamefont {Kurvits}}, \bibinfo {author}
  {\bibfnamefont {S.}~\bibnamefont {Ramanathan}}, \ and\ \bibinfo {author}
  {\bibfnamefont {R.}~\bibnamefont {Zia}},\ }\href@noop {} {\bibfield
  {journal} {\bibinfo  {journal} {Nature communications}\ }\textbf {\bibinfo
  {volume} {6}},\ \bibinfo {pages} {8636} (\bibinfo {year} {2015})}\BibitemShut
  {NoStop}%
\bibitem [{\citenamefont {Wang}\ \emph {et~al.}(2016)\citenamefont {Wang},
  \citenamefont {Rogers}, \citenamefont {Gholipour}, \citenamefont {Wang},
  \citenamefont {Yuan}, \citenamefont {Teng},\ and\ \citenamefont
  {Zheludev}}]{wang2016optically}%
  \BibitemOpen
  \bibfield  {author} {\bibinfo {author} {\bibfnamefont {Q.}~\bibnamefont
  {Wang}}, \bibinfo {author} {\bibfnamefont {E.~T.}\ \bibnamefont {Rogers}},
  \bibinfo {author} {\bibfnamefont {B.}~\bibnamefont {Gholipour}}, \bibinfo
  {author} {\bibfnamefont {C.-M.}\ \bibnamefont {Wang}}, \bibinfo {author}
  {\bibfnamefont {G.}~\bibnamefont {Yuan}}, \bibinfo {author} {\bibfnamefont
  {J.}~\bibnamefont {Teng}}, \ and\ \bibinfo {author} {\bibfnamefont {N.~I.}\
  \bibnamefont {Zheludev}},\ }\href@noop {} {\bibfield  {journal} {\bibinfo
  {journal} {Nature Photonics}\ }\textbf {\bibinfo {volume} {10}},\ \bibinfo
  {pages} {60} (\bibinfo {year} {2016})}\BibitemShut {NoStop}%
\bibitem [{\citenamefont {R{\'\i}os}\ \emph {et~al.}(2015)\citenamefont
  {R{\'\i}os}, \citenamefont {Stegmaier}, \citenamefont {Hosseini},
  \citenamefont {Wang}, \citenamefont {Scherer}, \citenamefont {Wright},
  \citenamefont {Bhaskaran},\ and\ \citenamefont
  {Pernice}}]{rios2015integrated}%
  \BibitemOpen
  \bibfield  {author} {\bibinfo {author} {\bibfnamefont {C.}~\bibnamefont
  {R{\'\i}os}}, \bibinfo {author} {\bibfnamefont {M.}~\bibnamefont
  {Stegmaier}}, \bibinfo {author} {\bibfnamefont {P.}~\bibnamefont {Hosseini}},
  \bibinfo {author} {\bibfnamefont {D.}~\bibnamefont {Wang}}, \bibinfo {author}
  {\bibfnamefont {T.}~\bibnamefont {Scherer}}, \bibinfo {author} {\bibfnamefont
  {C.~D.}\ \bibnamefont {Wright}}, \bibinfo {author} {\bibfnamefont
  {H.}~\bibnamefont {Bhaskaran}}, \ and\ \bibinfo {author} {\bibfnamefont
  {W.~H.}\ \bibnamefont {Pernice}},\ }\href@noop {} {\bibfield  {journal}
  {\bibinfo  {journal} {Nature Photonics}\ }\textbf {\bibinfo {volume} {9}},\
  \bibinfo {pages} {725} (\bibinfo {year} {2015})}\BibitemShut {NoStop}%
\bibitem [{\citenamefont {Koonen}(2018)}]{Koonen2018}%
  \BibitemOpen
  \bibfield  {author} {\bibinfo {author} {\bibfnamefont {T.}~\bibnamefont
  {Koonen}},\ }\href {http://jlt.osa.org/abstract.cfm?URI=jlt-36-8-1459}
  {\bibfield  {journal} {\bibinfo  {journal} {J. Lightwave Technol.}\ }\textbf
  {\bibinfo {volume} {36}},\ \bibinfo {pages} {1459} (\bibinfo {year}
  {2018})}\BibitemShut {NoStop}%
\end{thebibliography}%




\end{document}